\documentclass{article}

\usepackage{amssymb,amsmath,times}

\usepackage{anysize}

\begin{document}
\title{Linear differential equations to solve nonlinear
mechanical problems: A novel approach}

\author{C.~Radhakrishnan Nair\\ Institute of Mathematical
Sciences, Pangappara,\\ Thiruvananthapuram 695581, India\\
E-mail:~{\tt nairrcc@rediffmail.com}}

\maketitle

\begin{abstract}
 Often a non-linear mechanical problem is formulated as a non-linear
differential equation. A new method is introduced to find out new
solutions of non-linear differential equations if one of the solutions
of a given non-linear differential equation is known. Using the known
solution of the non-linear differential equation, linear differential
equations are set up. The solutions of these linear differential
equations are found using standard techniques. Then the solutions of
the linear differential equations are put into non-linear
differential equations and checked whether these solutions are also
solutions of the original non-linear differential equation. It is found
that many solutions of the linear differential equations are also
solutions of the original non-linear differential equation.
\end{abstract}

\section{Introduction}
Nonlinear mechanics is often said \cite{1} to be the last and third major
development of physics in the 20th century. The other two are
quantum mechanics and relativity. One of the major tools for
studying nonlinear mechanics is nonlinear differential
equations \cite{2,3}. Nonlinear differential equations appear prominently
in the study of fluid dynamics, cooperative and nondissipative
phenomena \cite{1,5,6}, General theory of relativity \cite{7} etc. Probably one of the
best ways to obtain maximum information about a nonlinear
mechanical system is to set up a nonlinear differential equation
and to find out the maximum number of explicit solutions of the
nonlinear differential equation. The larger the number of the
explicit solutions we know, the better our understanding of the
mechanical system, represented by the nonlinear differential
equations. Thus to find out the maximum number of solutions of a
given nonlinear differential equation is of utmost importance.
But to find out new explicit solutions of nonlinear differential
equations is often extremely difficult. For the last three
hundred years, mathematicians, physicists and engineers were
mostly concerned with the solutions of linear differential
equations. As a result we now know far more about solving linear
differential equations than solving nonlinear differential
equations. Here we show how a knowledge of solving a linear
differential equation can help us to locate new solutions of a
given nonlinear differential equation if one of the solutions of
the nonlinear differential equation is already known.

In the next paragraph, we shall discuss the basic properties of
the differential equations and their solutions that led to the
present work.

It is common knowledge that many differential equations have
some common solutions. Examples are in plenty. The harmonic
oscillator equation
\begin{equation}\label{eqn1}
\frac{d^2x}{dt^2}+x=0
\end{equation}
has two linearly independent solutions
\begin{align}
x &= \cos t\mbox{ and }\label{eqn2}\\
x &= \sin t\label{eqn3}
\end{align}
The sixth order linear differential equation
\begin{equation}\label{eqn4}
\frac{d^6x}{dt^6}+x=0
\end{equation}
has also \(x=\cos t\) and \(x=\sin t\) as solutions. Again the
second degree nonlinear differential equation
\begin{equation}\label{eqn5}
\left(\frac{dx}{dt}\right)^2+x^2-1=0
\end{equation}
has \(x=\cos t\) and \(x=\sin t\) as solutions. Therefore, it
seems reasonable to conclude that there is something common for
the three systems represented by the equations (\ref{eqn1}),
(\ref{eqn4}) and (\ref{eqn5}). What is this something that is
common?

An explicit solution of a differential equation is a function. A
function has a set of symmetries associated with it. For
example, \(\sin x\) is an odd function with a set of symmetries
associated with it. Thus when a linear differential equation and
a nonlinear differential equation have one common solution, the
two different differential equations represent two different
systems having one common set of symmetries. This idea of common
set of symmetries can be extended to any number of common
solutions for linear differential equations or for any set of
nonlinear differential equations. Warning! The symmetries
associated with differential equations can be altered by initial
and boundary conditions of the problem.

What we have just discussed about the ordinary differential
equations and their solutions is also true of partial
differential equations. The well known K dV equation
\begin{equation}\label{eqn6}
u_t-6uu_x+u_{xxx}=0
\end{equation}
has \def\sech{\mathop{\rm sech}\nolimits}
\begin{equation}\label{eqn7}
u(x,t)=-\frac{c}{2}\sech^2\left\{\frac{c^{1/2}}{2}(x-ct-x_0)\right\}
\end{equation}
as one of its solutions \cite{8}. Equation~(\ref{eqn7}) is also the
solution of the well known second order wave equation
\begin{equation}\label{eqn8}
u_{xx}=\frac{1}{c^2}u_{tt}
\end{equation}
Equation~(\ref{eqn7}) is also the solution of many other wave
equations. To cite one more example, the nonlinear Burgers'
equation
\begin{equation}\label{eqn9}
u_t+uu_{x}=0
\end{equation}
has \(u=\frac{x}{t}\) as one of its explicit solutions. The
linear partial differential equations
\begin{align}
&xu_x+tu_t=0\label{eqn10}\\
&2xu_x+t^2u_t=0\label{eqn11}
\end{align}
have also \(u=\frac xt\) as their solution.

We have earlier remarked that common solutions for different
differential equations imply that the corresponding differential
equations have common symmetry properties. This aspect is of
fundamental importance from the point of view of the theory of
differential equations. But here we show how the commonness of
some solutions of a linear differential equation and a nonlinear
differential equation can be exploited to find new solutions of
nonlinear differential equations.

\section{Illustration}
We shall illustrate our method of finding new solutions for both
ordinary nonlinear differential equations and nonlinear partial
differential equations.

We shall start with the ordinary nonlinear differential equation
\begin{equation}\label{eqn12}
\left(\frac{dx}{dt}\right)^4-x^4+2x^2-1=0
\end{equation}
We shall take \(x=\cos t\) as the already known solutions of
(\ref{eqn12}). We shall call this solution as the seed solution.
By differentiating the seed solution four times, we get the
fourth order linear differential equation
\begin{equation}\label{eqn13}
\frac{d^4x}{dt^4}-x=0
\end{equation}
By using the standard techniques of solving linear differential
equations, we can obtain three other solutions of (\ref{eqn13}).
They are
\begin{align}
x&=\sin t\tag{3}\\
x&=\cosh t \mbox{ and}\label{eqn14}\\
x&=\sinh t\label{eqn15}
\end{align}
The solutions (\ref{eqn3}), (\ref{eqn14}) and (\ref{eqn15}) are
solutions of a linear differential equation generated from one
of the solutions of the nonlinear differential equation.
Therefore, since the seed is the same for both linear and
nonlinear differential equations, there is a possibility that
some or all the solutions of the linear equation~(\ref{eqn13})
will also be the solutions of the nonlinear differential
equation~(\ref{eqn12}).

Substitution shows that \(x=\sin t\) and \(x=\cosh t\) are also
solutions of the nonlinear equation~(\ref{eqn12}). In a nut
shell, by setting up the fourth order linear differential
equation \(\frac{d^4x}{dt^4}-x=0\) from the known solution
\(x=\cos t\) of the nonlinear differential equation
\(\left(\frac{dx}{dt}\right)^4-x^2+2x^2-1=0\), we have succeeded in obtaining
two new solutions, namely \(x=\sin t\) and \(x=\cosh t\) of the
nonlinear differential equation.

>From the known solution \(x=\cos t\) of the nonlinear
differential equation~(\ref{eqn12}), we can also set up the
linear differential equation,
\begin{equation}
\frac{d^2x}{dt^2}+x=0\tag{1}
\end{equation}
As we have stated earlier, this equation has two linearly
independent solutions \(x=\cos t\) and \(x=\sin t\). Therefore,
if we had set up the second order linear differential
equation~(\ref{eqn1}), instead of the fourth order
equation~(\ref{eqn13}), we would have got only one new solution
\(x=\sin t\) for the nonlinear differential equation
\(\left(\frac{dx}{dt}\right)^4-x^4+2x^2-1=0\). In general, the
higher the order of the linear differential equation set up, the
greater the probability for finding new solutions to the
original nonlinear differential equation.

Note that the linear superposition \(x=\cos t+\sin t\) of the
linearly independent solutions \(x=\cos t\) and \(x=\sin t\), is
a solution of the linear differential equation~(\ref{eqn13}),
but \(x=\cos t+\sin t\) is not a solution of the nonlinear
differential equation~(\ref{eqn12}).

Now we shall move to the realm of partial differential
equations. Let us consider the first order, second degree
nonlinear wave equation
\begin{equation}\label{eqn16}
u_x^2=\frac{k^2}{\omega ^2}u_t^2
\end{equation}
where

\(u=u(x,t)\) is the dependent variable

\(x,t\) -- independent variables

\(k,\omega \) -- parameters
\begin{equation}\label{eqn17}
u=\cos(kx-\omega t)
\end{equation}
is taken as the known seed solution from~(\ref{eqn16}). By
differentiating~(\ref{eqn17}) twice with respect to \(x\) and
\(t\), we can obtain the second order linear partial
differential equation
\begin{equation}\label{eqn18}
u_{xx}=\frac{1}{c^2}u_{tt}
\end{equation}
It can be easily found by standard methods that
\begin{align}
u&=\sin(kx-\omega t)\label{eqn19}\\
u&=\cos h(kx-\omega t)\label{eqn20}\\
u&=\sin h(kx-\omega t)\label{eqn21}\\
u&=\sin (kx+\omega t)\label{eqn22}\\
u&=\cos (kx+\omega t)\label{eqn23}\\
u&=\sin h(kx+\omega t)\label{eqn24}\\
u&=\cos h(kx+\omega t)\label{eqn25}
\end{align}
are solutions of the second order wave equation
\(u_{xx}=\frac{1}{c^2}u_{tt}\). Now we can verify whether any of
these solutions is also a solution of the original nonlinear
differential equation~(\ref{eqn16}). In fact all the
solutions~(\ref{eqn19}) to~(\ref{eqn25}) are also solutions of
the nonlinear partial differential equation~(\ref{eqn16}).

It is interesting to note that the linear superpositions
\begin{align}
u&=\sin(kx+\omega t)+\cos(kx+\omega t)\mbox{ and }\label{eqn26}\\
u&=\sin(kx+\omega t)+\cos(kx-\omega t)\nonumber
\end{align}
are solutions of the second order linear wave equation
\(u_{xx}=\frac{k^2}{\omega^2}u_{tt}^2\) where as only
\(u=\sin(kx+\omega t)+\cos k(x+\omega t)\) is a solution of the nonlinear
differential equation
\[
u_x^2=\frac{k^2}{\omega ^2}u_t^2
\]
\(u=\sin(kx+\omega t)+\cos(kx-\omega t)\) is not a solution of the above
nonlinear differential equation. This implies that the nonlinear
wave equation \(u_x^2=\frac{k^2}{\omega ^2}u_t^2\) can represent only
superposition of waves moving in the same direction.
This simple result is of immense importance in the theory of wave equations
and the application of wave theory to fluid dynamics.

New solutions to well known nonlinear partial differential
equations, such as K dV equation and Burgers' equation will be
reported somewhere else.

\section{Limitation}

The above method suffers from a serious limitation. Starting
from the solution of a nonlinear differential equation, a higher
order differential equation can be set up only if the
derivatives higher than the first derivative do exist. For
example, if for the solution of a given first order nonlinear
differential equation, if the derivatives higher than the first
derivative do not exist, then we can not set up linear
differential equations of second or higher order, starting from
that particular seed solution.

\section*{Acknowledgement}
I am very happy to acknowledge the help received from L.~A.~Ajith
in the preparation of the manuscript.

\end{document}